\documentclass[pra,twocolumn,preprintnumbers,amsmath,amssymb,nofootinbib]{revtex4}
\usepackage{amsmath}
\usepackage{amssymb}
\usepackage{slashed}
\usepackage{graphicx}
\usepackage{graphics}
\usepackage{epsfig}
\usepackage{psfrag}
\usepackage{hyperref}

\usepackage{color}
\definecolor{comment}{rgb}{0.9,0,0}
\definecolor{holger}{rgb}{0,0.5,0.7}

\newcommand{\Eqref}[1]{Eq.~\eqref{#1}}
\newcommand{\hT}{h^{\text{T}}}

\begin{document}

\title{Asymptotically free scalar curvature-ghost coupling in Quantum Einstein
  Gravity} \date{\today} \author{Astrid Eichhorn, Holger Gies and Michael
  M. Scherer}

\pacs{}

\affiliation{\mbox{\it Theoretisch-Physikalisches Institut, Friedrich-Schiller-Universit{\"a}t Jena,}
\mbox{\it Max-Wien-Platz 1, D-07743 Jena, Germany}
\mbox{\it E-mail: {astrid.eichhorn@uni-jena.de, holger.gies@uni-jena.de,}}
\mbox{\it {michael.scherer@uni-jena.de}}
}

\begin{abstract} 
We consider the asymptotic-safety scenario for quantum gravity which
constructs a non-perturbatively renormalisable quantum gravity theory with the
help of the functional renormalisation group. We verify the existence of a
non-Gau\ss ian fixed point and include a running curvature-ghost coupling as a
first step towards the flow of the ghost sector of the theory. We find that
the scalar curvature-ghost coupling is asymptotically free and RG relevant in
the ultraviolet.  Most importantly, the property of asymptotic safety
discovered so far within the Einstein-Hilbert truncation and beyond remains
stable under the inclusion of the ghost flow.
\end{abstract}

\maketitle

\section{Introduction}
The construction of an internally consistent and falsifiable theory of quantum
gravity is one of the major challenges of modern theoretical physics. As the
perturbative quantisation of the Einstein-Hilbert action yields a
non-renormalisable theory
\cite{'tHooft:1974bx,Goroff:1985sz,vandeVen:1991gw,Christensen:1979iy} several
alternative approaches have been proposed: A change in the degrees of freedom
and of the microscopic action, as well as a different approach to
quantisation, or assumptions about a discrete nature of spacetime offer
possible routes to a predictive theory of quantum gravity. However, the
possibility remains that the apparent nonrenormalisability is not a failure of
Einstein gravity but rather of the simple perturbative quantisation
scheme. This is the underlying viewpoint of, e.g., lattice simulations of the
gravitational path integral
\cite{Ambjorn:2005db,Ambjorn:2004pw,Ambjorn:2004qm}.

Within analytical continuum approaches to a non-perturbative quantisation of
gravity, Weinberg's asymptotic-safety scenario
\cite{Weinberg:1980gg,Weinberg:1996kw,Weinberg:2009ca} represents a possible way for a predictive theory of quantum gravity: Weinberg
argued that, if the non-perturbative renormalisation group trajectory of a
quantum field theory approaches a non-Gau\ss{}ian fixed point (NGFP) in the
ultraviolet (UV), the UV limit can safely be taken. If furthermore the NGFP
has a finite number of relevant directions, the theory has predictive power. If
moreover an RG trajectory emanating from the NGFP can be connected
continuously with a regime in the infrared, where both the cosmological
constant and Newton's constant are positive and small, the asymptotic-safety
scenario is consistent with our universe as we observe it now.

A suitable tool to study this scenario is the functional renormalisation group
(RG). Formulated in terms of the Wetterich equation
\cite{Wetterich:1992yh,Wetterich:1989xg}, the functional RG describes the change of the
effective average action $\Gamma_k$ as a function of a momentum scale $k$:
\begin{equation}
 \partial_t\Gamma_k
        =\frac{1}{2}\mathrm{STr}\{[\Gamma^{(2)}_k+R_k]^{-1}(\partial_tR_k)\},
        \quad \partial_t=k\frac{d}{dk}. \label{eq:wetterich}
\end{equation}
Here, $\Gamma_k^{(2)}$ denotes the second functional derivative of this
effective action with respect to the fields, which is the full inverse
propagator at the scale $k$. The quantity $R_k$ is an infrared regulator, which suppresses
contributions of modes with momenta $p^2 <k^2$ to the supertrace.
Projecting the equation onto a given operator on the left- and right-hand side
yields the non-perturbative $\beta$ function of the coupling associated with
this operator. 

The functional RG for gravity has been pioneered by Reuter
\cite{Reuter:1996cp}. It proceeds in many ways similar to the corresponding
quantisation of Yang-Mills theories, see, e.g.,
\cite{Reuter:1993kw,Reuter:1994zn,Reuter:1997gx,Pawlowski:2005xe,Gies:2006wv}. Whereas
various systematic and consistent nonperturbative approximation schemes, such
as derivative or vertex expansions have been devised and successfully applied
in a large variety of cases ranging from critical phenomena to strong-coupling
problems in gauge theories, the choice of truncations of the effective action
in gravity has partly been guided by the available computational tools, most
notably the heat-kernel expansion.

Early calculations have focussed on the Einstein-Hilbert truncation where the
existence of a NGFP has been observed for the first time
\cite{Reuter:1996cp,Souma:1999at}. The fixed point has remained remarkably
stable upon the inclusion of higher orders of the curvature
\cite{Lauscher:2001ya,Reuter:2001ag,Litim:2003vp,Granda:1998wn,Lauscher:2001rz,%
Lauscher:2002sq,Machado:2007ea,Codello:2008vh,Forgacs:2002hz,Niedermaier:2002eq,%
Niedermaier:2003fz},
even up to $R^{6}$ \cite{Codello:2007bd}. Recently, new techniques have
allowed to go beyond the class of $f(R)$ truncations and include, e.g., the
operator $C_{\mu\nu\kappa\lambda}C^{\mu\nu\kappa\lambda}$ involving the Weyl
tensor among the gravitational interactions \cite{Benedetti:2009rx}. Not only
the existence of the non-Gau\ss ian fixed point has been confirmed in a
variety of studies, but also the critical exponents classifying the relevant
directions at the fixed point show a satisfactory convergence upon increasing
the truncation. All these investigations have accumulated a substantial body
of evidence for asymptotic safety of Quantum Einstein Gravity (QEG) (for
reviews see \cite{Niedermaier:2006wt, Litim:2008tt, Percacci:2007sz}). This
stimulated, of course, also the study of phenomenological implications of the
existence of a fixed point (see e.g.  \cite{Manrique:2009tj} for references).

As is known from certain gauges in Yang-Mills theories, e.g., the Landau gauge
or the Coulomb gauge, ghosts can play an important if not dominant role in the
strong-coupling regime of a gauge theory
\cite{Kugo:1979gm,Gribov:1977wm,Zwanziger:2003cf,Alkofer:2000wg,Pawlowski:2005xe,%
  Fischer:2006ub}. This is because the strong-coupling regime can be
entropically dominated by field configurations near the Gribov horizon which
induce an enhancement of the ghost propagator. As the Gribov ambiguity is also
present in standard gauges in general relativity \cite{Das:1978gk}, similar
mechanisms may become relevant in QEG.
However, investigations of the non-Gau\ss ian fixed point
in QEG have neglected running couplings in the ghost sector so far, mainly
for technical reasons. Only a classical ghost term has been considered. In
this work, we take a first step in this direction and examine the RG flow in
the ghost sector. More specifically, we truncate the theory space of QEG down
to the following action:
\begin{eqnarray}
 \Gamma_k&=&\Gamma_{\mathrm{EH}}+ \Gamma_{\mathrm{gf}}+\Gamma_{\mathrm{gh}}+
 \Gamma_{\mathrm{R\,gh}},\label{truncation}
\end{eqnarray}
where 
\begin{eqnarray}
\Gamma_{\mathrm{EH}}&=& 2 \bar{\kappa}^2 Z_{\text{N}} (k)\int 
d^d x \sqrt{\gamma}(-R+ 2 \bar{\lambda}(k))\label{eq:GEH},\\
\Gamma_{\mathrm{gf}}&=& \frac{Z_{\text{N}}(k)}{2\alpha}\int d^d x
\sqrt{\bar g}\, \bar{g}^{\mu \nu}F_{\mu}[\bar{g}, h]F_{\nu}[\bar{g},h]\label{eq:Ggf},\\ 
\Gamma_{\mathrm{gh}}&=& - \sqrt{2}\,Z_{\mathrm{c}}(k)\int d^d x \, \sqrt{\bar{g}}\, \bar{c}_{\mu}
      \mathcal{M}^{\mu}{}_{\nu}c^{\nu}\label{eq:Ggh},\\ 
\Gamma_{\mathrm{R\,gh}}&=& \bar\zeta (k) \int d^d x \sqrt{\gamma}\, \bar{c}^{\mu}R c_{\mu}, 
\label{eq:GRgh}
\end{eqnarray}
and $\bar{\kappa}^2 = \frac{1}{32 \pi G_{\text{N}}}$. The Einstein-Hilbert
action $\Gamma_{\text{EH}}$ contains the dimensionful Newton constant
$G_{\text{N}}$, the running graviton wave function renormalisation
$Z_{\text{N}}(k)$, and the running cosmological constant
$\bar{\lambda}(k)$. For the necessary gauge fixing of metric fluctuations, we
apply the background-field method
\cite{Abbott:1980hw,Reuter:1994zn,Reuter:1996cp,Freire:2000bq}, where the full
metric
\begin{equation}
 \gamma^{\mu \nu}= \bar{g}^{\mu \nu}+ h^{\mu \nu},
\end{equation}
is decomposed into the background metric $\bar{g}^{\mu \nu}$ with its
compatible covariant derivative $\bar{D}_{\lambda}$. $h^{\mu \nu}$ denotes the
fluctuations around this background which do not have to obey any constraints
concerning their amplitude. This is a crucial difference to a perturbative
treatment on a fixed background. The curvature scalar constructed from the
full metric is denoted by $R$, the one pertaining to the background metric by
$\bar{R}$.  The gauge-fixing action $\Gamma_{\text{gf}}$ contains the
gauge-fixing condition $F_\mu[\bar{g},h]$, which reads for the 
background-covariant generalisation of the harmonic gauge: 
\begin{equation}
 F_{\mu}[\bar{g}, h]= \sqrt{2} \bar{\kappa} \left(\bar{D}^{\nu}h_{\mu
   \nu}-\frac{1+\rho}{d}\bar{D}_{\mu}h^{\nu}{}_{\nu} \right). 
\end{equation}
In \Eqref{eq:Ggf}, we ignore any independent running of
$\alpha(k)\to\alpha=$const. The appropriate ghost term $\Gamma_{\mathrm{gh}}$
contains the Faddeev-Popov operator
\begin{eqnarray}
 {\mathcal{M}^{\mu}{}_{\nu}}&=& \bar{g}^{\mu \rho}\bar{g}^{\sigma \lambda}\bar{D}_{\lambda}(\gamma_{\rho \nu}D_{\sigma}+ \gamma_{\sigma \nu}D_{\rho})\\
&{}&- 2 \frac{1+\rho}{d}\bar{g}^{\rho \sigma}\bar{g}^{\mu \lambda}\bar{D}_{\lambda}\gamma_{\sigma \nu}D_{\rho},\nonumber
\end{eqnarray}
and $Z_{\mathrm{c}} (k)$ denotes a wave function renormalisation for the ghosts, which we
will later set to $Z_{\mathrm{c}}(k)=1$ in our calculations; this amounts to a classical
treatment of the ghost "kinetic" term. The $\Gamma_{\mathrm{R}\,\mathrm{gh}}$ term in \Eqref{eq:GRgh}
contains the running curvature-ghost coupling $\bar\zeta (k)$ which is of
central interest in this work. In any dimension, it corresponds to a
  marginal coupling in a perturbative power-counting classification, as the
  ghosts carry canonical dimension $\frac{d-2}{2}$. Since such a
  classification does no longer hold at a non-Gau\ss ian fixed point, it is a
  crucial question whether this curvature-ghost coupling becomes
  relevant or irrelevant because of the interactions. 

In the remainder of this work, we drop the argument $k$ of the couplings which
are implicitly understood as running couplings.

\section{Computational method}

Our truncation \eqref{truncation} defines a hypersurface in theory space. The
solution of the flow equation \eqref{eq:wetterich} provides us with an RG
trajectory in this hypersurface, once an initial condition is specified. For
unspecified initial conditions, the flow equation defines a vector field on
this hypersurface. In order to arrive at an explicit representation of this
vector field in terms of the flow of the couplings, we need to project the
right-hand side of the flow equation onto the operators of our truncation.

For this, we perform the computation on a maximally symmetric background
metric of a $d$ dimensional sphere of radius $r$ in Euclidean space where
\begin{eqnarray}
\bar{R}&=& \frac{d(d-1)}{r^2}, \quad \bar{R}_{\mu \nu}=\bar{g}_{\mu \nu}\frac{\bar{R}}{d},\nonumber\\
\quad\int d^d x
\sqrt{\bar{g}}&=& \frac{\Gamma(d/2)}{\Gamma(d)}(4 \pi r^2)^{\frac{d}{2}}.
\label{eq:dspheres}
\end{eqnarray}
Whereas this leads to an enormous simplification of the calculations, this
background does not allow to disentangle the flow of $\Gamma_{\mathrm{R}\,\text{gh}}$
from that of, e.g., an operator of the form $\int d^dx \sqrt{\gamma}\, \bar{c}^\mu
R_{\mu\nu} c^{\nu}$. Moreover, as we set $\gamma^{\mu\nu}$=$\bar g^{\mu\nu}$
in the flow equation, an operator of the form $ \int d^d x \sqrt{\bar{g}}
\bar{c}^{\mu}{R}c_{\mu}$ can not be disentangled from $\Gamma_{\mathrm{R}\,\text{gh}}$.
Our calculation shares these ambiguities with most of the other works on the
asymptotic-safety scenario for QEG, as the derivation of the flow equation
with two distinct metrics and more complex backgrounds is highly
non-trivial.

A simplifying but sufficiently unique choice for the background ghost fields
is given by covariantly constant ghosts,
\begin{equation}
 \bar{D}^{\mu}c^{\lambda}=0,\label{eq:covghost}
\end{equation}
which allows to disentangle the flow of $\bar\zeta$ from that of the ghost
wave function renormalisation $Z_{\mathrm{c}}$.

Past works in QEG evaluated the trace on the right-hand side of the flow
equation by invoking a propertime representation and using heat-kernel
techniques. For our purposes, we explicitly invert $\Gamma^{(2)}_k$ on a $d$
sphere to get the graviton propagator in terms of a basis of hyperspherical
harmonics. The trace operation over the eigenvalues can then be evaluated
explicitly.

This technique has the advantage that it allows for a straightforward
inclusion of external ghost fields, and hence the flow of the coupling
parameter of the curvature-ghost term $\bar\zeta$ is accessible with our
method. For this, we use a decomposition of $\Gamma^{(2)}_k+R_k$ into an
inverse propagator matrix contribution $\mathcal{P}_k=\Gamma_k^{(2)}[\bar{c}=0=c]+R_k$
containing the regulator but no external ghost fields and a fluctuation matrix
contribution $\mathcal{F}$ containing external ghost fields. The components of
$\mathcal{F}$ are either linear or bilinear in the ghost fields. With this
decomposition, the right-hand side of the flow equation can be expanded as
follows:
\begin{eqnarray}
 \partial_t \Gamma_k&=& \frac{1}{2}{\rm STr} \{
 [\Gamma_k^{(2)}+R_k]^{-1}(\partial_t R_k)\}\label{eq:flowexp}\\
&=&\frac{1}{2} {\rm STr}\tilde{\partial}_t \ln (\mathcal{P}_k+\mathcal{F})
\nonumber\\
&=& \frac{1}{2} {\rm STr} \tilde{\partial}_t\ln \mathcal{P}_k
+\frac{1}{2}\sum_{n=1}^{\infty}\frac{(-1)^{n-1}}{n} {\rm
  STr} \tilde{\partial}_t(\mathcal{P}_k^{-1}\mathcal{F})^n,\nonumber
\end{eqnarray}
where the derivative $\tilde{\partial}_t$ in the second line by defintion acts
only on the $k$ dependence of the regulator, $\tilde{\partial}_t=
\partial_t R_k\frac{d}{dR_k}$.

On the background of $d$ spheres, the left-hand side of the
  flow equation for the scalar curvature-ghost coupling after setting
  $\gamma^{\mu\nu}=\bar{g}^{\mu\nu}$ takes the form
\begin{eqnarray}
 \partial_t \Gamma_{\mathrm{R\,gh}}&=& (\partial_t \bar\zeta) \bar{c}^{\mu}
 c_{\mu} R \int d^d x \sqrt{\bar{g}} \label{eq:zetaproject}\\
&=& (\partial_t \bar\zeta) \bar{c}^{\mu} c_{\mu}  
\frac{\Gamma(d/2)}{\Gamma(d)}\, d(d-1)\, (4\pi)^{d/2}\, (r^2)^{\frac{d}{2}-1}.
\nonumber
\end{eqnarray}
As $\mathcal{F}$ contains terms linear and bilinear in the ghosts, the flow of
$\bar\zeta$ is fully determined by  the $n=1$ and $n=2$ terms of the expansion
of the flow equation \Eqref{eq:flowexp}.

We evaluate the fluctuation matrix entries in the Landau-DeWitt gauge $\alpha=0$, as this is a fixed point of the RG
flow \cite{Ellwanger:1995qf,Litim:1998qi}, and we set the second gauge
parameter $\rho=1$ for our evaluation of the Einstein-Hilbert sector; see
App.\ref{EH}. Then we apply the York decomposition \cite{York:1973ia} to the
graviton $h_{\mu \nu}$ to identify its trace part $h = \bar{g}^{\mu
  \nu}h_{\mu \nu}$, its transverse traceless part $\hT_{\mu \nu}$, and the
traceless longitudinal vector and scalar parts $\hat\xi_\mu$ and $\hat\sigma$:
\begin{equation} 
 h_{\mu \nu} = \hT_{\mu \nu} + \bar{D}_{\mu }\hat{\xi}_{\nu}+
 \bar{D}_{\nu}\hat{\xi}_{\mu}+ \bar{D}_{\mu}\bar{D}_{\nu}\hat{\sigma}-
 \frac{1}{d}\bar{g}_{\mu \nu}\bar{D}^2 \hat{\sigma} +\frac{1}{d}\bar{g}_{\mu
   \nu}h \label{York}. 
\end{equation}
In the Landau-DeWitt gauge, the contribution of all graviton modes except for the
transverse traceless tensor mode is zero, as can be seen by considering the
generic form of the flow equation: The full propagator (including a regulator)
of vector, scalar and conformal mode is $\propto \alpha$ in the background
version of the harmonic gauge \Eqref{Gamma2}. Thereby, each term in the
$\mathcal{P}_k^{-1}\mathcal{F}$ expansion containing the vector, scalar, or
conformal mode will always be $\propto \alpha$, too.

 Hence, we arrive at the five diagrams displayed in Fig. \ref{diagrams}, in which only the transverse
 traceless graviton mode propagates:
 \begin{figure}[!here]
  \includegraphics[scale=0.45]{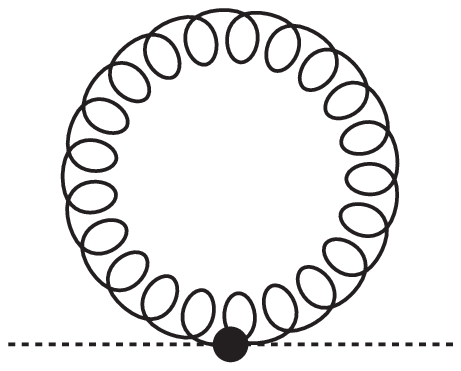}
  \includegraphics[scale=1.2]{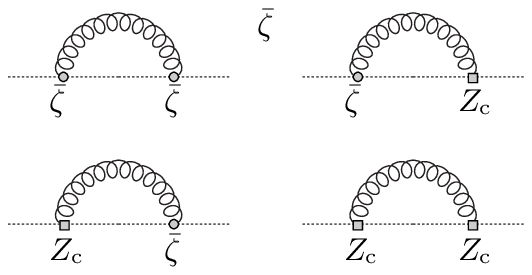}
  \caption{Single diagram on the top: tadpole contribution to the running of
    $\bar\zeta$. Four diagrams on the bottom: Two-vertex contribution to the
    wave function renormalisation of the ghost and the curvature-ghost coupling $\bar{\zeta}$.  Spiral
    lines denote gravitons, dashed lines denote ghosts. The regulator insertion
   in the internal propagators is implicitly understood.} 
\label{diagrams}
 \end{figure}
The two-vertex diagram appears in four different versions, as two different
antighost-ghost-graviton vertices exist in our truncation, namely one $\propto
\bar\zeta$ and one $\propto Z_{\mathrm{c}}$. Accordingly, there is a diagram with two
$\bar\zeta$ vertices, one with two $Z_{\mathrm{c}}$ vertices and two mixed ones. Now we can
make use of the fact that the equation of motion for the transverse traceless
part of $h^{\mu\nu}$ vanishes identically on $d$ spheres for vanishing ghost fields,
\begin{eqnarray}
&{}&\left(\delta (\sqrt{g}R)\right)\vert_{\hT_{\mu \nu}}\\
&{}&= \left(- h^{\mu \nu}R_{\mu \nu}+ \bar{D}^{\mu}\bar{D}^{\nu}h_{\mu \nu}- \bar{D}^2 h+\frac{1}{2}h R\right)\vert_{\hT_{\mu \nu}}\nonumber\\
&{}&\stackrel{\text{$d$ sphere}}{=}0.\nonumber
\end{eqnarray}

Hence, the antighost-ghost-graviton vertex which is $\propto \bar\zeta$ does
not receive a contribution from the transverse traceless mode, when evaluated
on the Euclidean deSitter background. \footnote{This fact is obvious for maximally symmetric spaces, where $R_{\mu \nu} \sim  g_{\mu \nu}R$. Hence, a different choice of background would require the evaluation of some of the ghost self-energy diagrams. However, this does not  imply that the final result for the $\beta$ function is background  dependent, as the fluctuation matrix entering the tadpole diagram would also  change on a different background. A maximally symmetric background simply  corresponds to a very efficient organisation of the flow equation also in  terms of diagrams.} 
Thereby, three of the four sunset diagrams are
zero. Finally, the diagram with two $\propto Z_{\mathrm{c}}$ vertices vanishes for
covariantly constant ghost background fields, as the ghost kinetic term can be
brought into the following symbolic form by partial integration
\begin{equation}
 \Gamma_{\text{gh}} \sim \int d^d x \sqrt{\bar{g}}\left(\bar{D}\bar{c}\right)\left(D c
 \right).
\end{equation}
Upon deriving the antighost-ghost-graviton vertex connected to the antighost
from this equation, the result is always proportional to $\bar{D}\bar{c}$, and
hence the diagram vanishes on the covariantly constant ghost background,
cf. \Eqref{eq:covghost}. This simply reflects the fact that this diagram
contributes solely to the running of the ghost wave function renormalisation.

To summarise, only the tadpole diagram contributes to the running of
$\bar\zeta$, corresponding to the $n=1$ term in the expansion of the flow
equation \Eqref{eq:flowexp}.  To evaluate the tadpole diagram, the necessary
fluctuation matrix entry projected onto a $d$ sphere is 
\begin{equation}
\delta^{2} \Gamma_{R\,\mathrm{gh}}|_{\hT}= \bar\zeta \bar{c}^{\alpha}c_{\alpha} \, \hT{}^{\kappa \lambda }
\delta^{\mu}_{\kappa}\delta^{\nu}_{\lambda}
\bigg\{ -\frac{d(d-3)+4}{2 r^2}+\frac{\bar{D}^2}{2}\bigg\}\hT_{\mu \nu}.
\end{equation}
Next, we need the explicit form of the transverse traceless graviton
propagator. The second variation of the Einstein-Hilbert action including the
gauge fixing with respect to the metric takes the form \cite{Lauscher:2001ya}
\begin{eqnarray}
 \delta^2\Gamma_{\mathrm{EH+gf}} &=& 2 \bar{\kappa}^2 Z_{\text{N}} \int d^dx\,
 \sqrt{\bar{g}}
 h_{\mu \nu} \bigg\{\nonumber\\
&{}&-\Bigl(\frac{1}{2}\delta^{\mu}_{\rho}\delta^{\nu}_{\sigma}
+ \frac{1-2 \alpha}{4 \alpha}\bar{g}^{\mu \nu}\bar{g}_{\rho \sigma} \Bigr)
\bar{D}^2 \nonumber\\
&{}&+\frac{1}{4} (2 \delta^{\mu}_{\rho}\delta^{\nu}_{\sigma}
- \bar{g}^{\mu \nu}\bar{g}_{\rho \sigma})(\bar{R}-2 \bar{\lambda})\nonumber\\
&{}&+\bar{g}^{\mu \nu}\bar{R}_{\rho \sigma}-
\delta^{\mu}_{\sigma}\bar{R}^{\nu}{}_{\rho}
-
\bar{R}^\nu{}_\rho{}^\mu{}_\sigma 
 \nonumber\\
&{}&+ \frac{1-\alpha}{\alpha}(\bar{g}^{\mu \nu}\bar{D}_{\rho}\bar{D}_{\sigma} 
- \delta^{\mu}_{\sigma}\bar{D}^{\nu}\bar{D}_{\rho}) \bigg\}h^{\rho \sigma}
\nonumber\\
&=& \int d^d x \sqrt{\bar g}\, h^{\mu\nu}\,
(\Gamma^{(2)}_h)_{\mu\nu\kappa\lambda}\, h^{\kappa\lambda}. \label{Gamma2}
\end{eqnarray}
In the Landau-DeWitt gauge, we need only the inverse propagator for the transverse
traceless tensor on the $d$ sphere:
\begin{equation}
 \left(\Gamma_{\hT}^{(2)}\right)_{\mu \nu \kappa \lambda} =
\bar{g}_{\mu \nu}\bar{g}_{\kappa \lambda}\,\bar{\kappa}^2 Z_{\text{N}}
\left(\frac{d(d-3)+4}{r^2} - 2 \bar{\lambda}
-\bar{D}^2\right)
. \label{h_inv_prop}
\end{equation}
Now we can invert the two-point function to arrive at the full regularised
$k$-dependent propagator, 
\begin{equation}
\left(\Gamma_{\hT}^{(2)}+R_{k,\hT}\right)_{\mu \nu \kappa \lambda} 
G_{\ \ \ \rho \sigma }^{\kappa \lambda}
= \frac{1}{\sqrt{\bar{g}}}\delta^d(x-x')\bar{g}_{\mu \rho}\bar{g}_{\sigma \nu},
\label{hGreen},
\end{equation}
where symmetrisation of $\mu \leftrightarrow \nu$ and $\rho \leftrightarrow \sigma$ on the left-hand side is understood implicitly. 
At this stage, we take advantage of the existence of a basis for
traceless tensor functions on the $d$ sphere: the tensor hyperspherical
harmonics $T^{lm}_{\mu \nu}(x)$. They fulfil a completeness and an
orthogonality relation, and are eigenfunctions of the covariant Laplacean
$\bar{D}^2$,
\begin{eqnarray}
 \frac{\delta^d(x-x')}{\sqrt{\bar g}}\bar{g}_{\mu \rho}\bar{g}_{\nu \sigma}&=&
 \sum_{l=2}^{\infty}\sum_{m=1}^{D_l}T_{\mu \nu}^{lm}(x)T_{\rho
   \sigma}^{lm}(x'),\qquad\label{complete} \\
\delta^{lk}\delta^{mn}&=&\int d^d x \, \sqrt{\bar{g}}\, \bar{g}^{\mu
  \rho}\bar{g}^{\nu \sigma}T^{lm}_{\mu \nu}(x)T^{kn}_{\rho \sigma}(x), \label{tortho}\nonumber\\
 -\bar{D}^2 T_{\mu \nu}^{lm}(x)&=& \Lambda_lT_{\mu \nu}^{lm}(x).\label{tsh}
\end{eqnarray}
For the tensor mode there is a $D_l$-fold degeneracy of
the hyperspherical harmonics for fixed $l$ but different $m$
\cite{Rubin:1984tc, Rubin:1983be} where
\begin{equation}
D_l=
\frac{(d+1)(d-2)(l+d)(l-1)(2l+d-1)(l+d-3)!}{2(d-1)!(l+1)!}.
\end{equation}
The eigenvalues of the Laplacean are given by
\begin{equation}
\Lambda_l=\frac{l(l+d-1)-2}{r^2}.
\end{equation}
As the hyperspherical harmonics form a basis for functions on the $d$ sphere, we
can expand the Green's function as follows:
\begin{equation}
 G(x-x')_{\mu \nu \rho \sigma}= \sum_{l=2}^{\infty}\sum_{m=1}^{D_l}
a_{lm}\,T^{lm}_{\mu \nu}(x)T^{lm}_{\rho \sigma}(x')\label{greenh},
\end{equation}
with expansion coefficients $a_{lm}$.  We insert our expression
\Eqref{h_inv_prop} into \Eqref{hGreen}, and use the eigenvalue equation
\Eqref{tsh}. As the regulator is some function of $-\bar{D}^2$, it turns into
the same function of $\Lambda_l$ in the hyperspherical-harmonics basis.

Applying the completeness relation allows to rewrite the right-hand side of
the definition of the Green's function \Eqref{hGreen}. By a comparison of
coefficients with respect to the hyperspherical-harmonics basis, we obtain
\begin{eqnarray}
 a_{lm}&=&\bigg( \bar{\kappa}^2
 Z_{\text{N}}\Bigl(\frac{d(d-3)+4}{r^2}-2\bar{\lambda}
 +\Lambda_l\Bigr)+R_{k,l}\bigg)^{-1}\,,\label{glmh}
\end{eqnarray}
for $l \geq 2$. Here, we have assumed that the argument of the regulator $R_{k}(x)$ is a
function of the Laplacean, $x=x(\bar{D}^2)$, such that
\begin{equation}
R_{k,l}:=R_{k}\Big(x\big(\Lambda_l\big)\Big).
\end{equation}
From the expression \Eqref{glmh}, it is obvious that the cosmological constant
is similar to a wrong-sign mass term for the graviton modes.\newline The $n=1$ term
in the expansion of the flow \Eqref{eq:flowexp} finally reads
\begin{widetext}
\begin{eqnarray}
 {\rm Tr} \left(\mathcal{P}^{-1}\mathcal{F} \right)
= {\rm Tr} \Biggl(\sum_{l=2}\sum_{m=1}^{D_l}
\frac{\bar\zeta\,\bar{c}^{\alpha}c_{\alpha}\, T^{lm}_{\mu \nu}(x) T^{lm\, \kappa \lambda}(y)
 \delta^{\mu}_{\kappa}\delta^{\nu}_{\lambda}}
{\bar{\kappa}^2 Z_{\text{N}} \left( \frac{(d(d-3)+4)}{r^2}- 2 \bar{\lambda} 
+\Lambda_{l}\right)+ R_{k,l}}
\left\{-\frac{d(d-3)+4}{2 r^2}- \frac{\Lambda_l}{2} \right\}\Biggr),
\end{eqnarray}
where the trace implies an integration with measure $\int d^d x \sqrt{\bar{g}}
\int d^d y \sqrt{\bar{g}} \frac{\delta^d(x-y)}{\sqrt{\bar{g}}}$. Invoking the
orthogonality relation \Eqref{tortho} and evaluating the $\tilde\partial_t$
derivative, we end up with the following expression:
\begin{eqnarray}
 {\rm Tr} \tilde{\partial}_t\left(\mathcal{P}^{-1}\mathcal{F}\right)
=\sum_{l=2}\frac{-\bar\zeta\,\bar{c}^{\alpha} c_{\alpha}\, D_{l} 
\partial_t R_{k,l}}{\left(\bar{\kappa}^2 Z_{\text{N}} 
\left( \frac{(d(d-3)+4)}{r^2}- 2 \bar{\lambda} +\Lambda_{l}\right)
+ R_{k,l}\right)^2} \left\{-\frac{d(d-3)+4}{2r^2}-
\frac{\Lambda_l}{2}\right\}.
\label{eq:pattilde}
\end{eqnarray}
\end{widetext}
We parameterise the regulator function for the transverse traceless mode by
\begin{equation}
 R_{k}(x)= x\, \mathfrak r\left(\frac{x}{Z_{\text{N}} \bar{\kappa}^2 k^2}\right),
\end{equation}
where the shape function $\mathfrak r(y)$ specifies the details of the
Wilsonian momentum-shell integration. Different choices correspond to
different RG schemes. As we need to expand in the curvature radius $r$ in
order to project the flow onto the truncation, an analytic shape function is
required. Here, we work with an exponential shape function $\mathfrak
r(y)=\frac{1}{e^{y}-1}$ as an example. Moreover, the regulator can be adjusted
to the flow of the spectrum of the full propagator \cite{Gies:2002af}, by
choosing $y = \frac{\Gamma_{\hT}^{(2)}}{Z_{\text{N}} \bar{\kappa}^2 k^2}$ evaluated on the
background field. For the graviton, this yields
\begin{equation}
 \partial_t R_k(y)= -(2-\eta) y \,\mathfrak{r}'\, \Gamma_{\hT}^{(2)} 
+ (\mathfrak{r} + y \,\mathfrak r') \partial_t \Gamma_{\hT}^{(2)},
\end{equation}
where the prime denotes the derivative of $\mathfrak r(y)$ with respect to
$y$. In addition, we have introduced the graviton anomalous dimension 
\begin{equation}
\eta = -\partial_t \ln Z_{\text{N}}.\label{eq:eta}
\end{equation}
(Our choice of a spectrally adjusted regulator corresponds to type III in \cite{Codello:2008vh}.)
For the trace in \Eqref{eq:pattilde}, i.e., the sum over $l$, we invoke the
Euler-MacLaurin formula that transforms the sum over $l$ into an integral. We
expand the result in powers of the $d$-sphere curvature $r$ amd project the
result onto the power of $(r^2)^{\frac{d}{2}-1}$ in order to perform a
comparison of coefficients with respect to the scalar curvature-ghost term,
cf. \Eqref{eq:zetaproject}.  
Incidentally, all non-integral terms in the Euler-MacLaurin formula do
not contribute to the scalar curvature-ghost coupling. 

Our method also applies straightforwardly to the Einstein-Hilbert sector,
where our results confirm the asymptotic-safety scenario of QEG obtained in
other gauges and with other regulators, see App. \ref{EH}.  As we use an
unprecedented combination of the Landau-DeWitt gauge $\alpha=0$ together with a spectrally
adjusted regulator, our results in this sector represent an independent
confirmation of asymptotic safety.

\section{Results}

For the discussion of the fixed-point structure of QEG, we introduce
dimensionless renormalised couplings $G$, $\lambda$, and $\zeta$ which are
related to the bare quantities by
\begin{eqnarray}
G&=&\frac{G_{\text{N}}}{Z_{\text{N}} k^{2-d}}
=  \frac{1}{32\,  \pi\,  \bar{\kappa}^2\,  Z_{\text{N}}\, k^{2-d}},\nonumber\\
%
\lambda&=& \bar{\lambda} k^{-2}, \quad \Rightarrow \partial_t \lambda = -2
\lambda + k^{-2}\partial_t \bar{\lambda},\nonumber\\
\zeta&=& \bar\zeta/Z_{\mathrm{c}}.
\end{eqnarray}
The running of the wave function renormalisations of graviton and ghost,
$Z_{\text{N}}$ and $Z_{\mathrm{c}}$, respectively, are governed by the corresponding
anomalous dimensions, $\eta$ (see \Eqref{eq:eta}), and 
%
$\eta_{\mathrm{c}} = -\partial_t \ln Z_{\mathrm{c}}$.
%

\begin{figure}[!here]
 \includegraphics[scale=0.4]{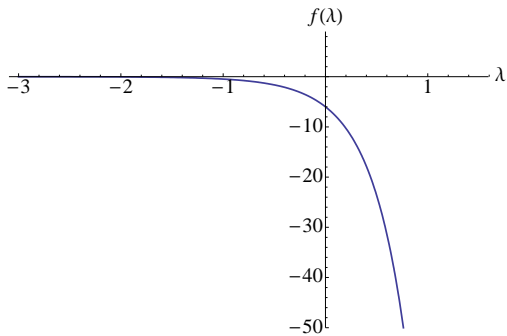}
\caption{$f(\lambda)$ is negative for all values of $\lambda$ and
  asymptotically tends to zero for $\lambda \rightarrow - \infty$, whereas it
  is unbounded for positive $\lambda$.}
\label{fig:f}
\end{figure}

From this point on, we confine ourselves to $d=4$, even though the
calculations in the remainder are straightforwardly generalisable to $d\neq
4$, see also \cite{Fischer:2006at} \cite{Fischer:2006fz}. 
%
%
%
With these prerequisites, we can now state our result for the graviton-tadpole
induced flow of the coupling $\zeta$:
\begin{eqnarray}
 \partial_t \zeta &=& \eta_{c} \zeta +\frac{25 G\zeta}{96\pi}
 \Bigl\{(e^{4 \lambda}-2 e^{2 \lambda})\left(2 \lambda+ \partial_t
 \lambda-\frac{\eta}{4}\right)-e^{2 \lambda}\nonumber\\ 
&+&\left((4\lambda-1)\left(\frac{\lambda}{2}+\partial_t
 \lambda\right)-\frac{\lambda\eta}{4}\right)\bigl({\rm Ei}(2 \lambda)- {\rm
   Ei}(4 \lambda)\bigl)\Bigr\}.\nonumber \\
&&\label{eq:zetares}
\end{eqnarray}
This flow equation has a Gau\ss{}ian fixed point $\zeta_\ast=0$. Let us
investigate the properties of this fixed point in the ghost sector for the
case that the remaining system is at the non-Gau\ss ian fixed-point of the
Einstein-Hilbert sector, $G\to G_\ast$, $\lambda\to\lambda_\ast$. For this, we
evaluate \Eqref{eq:zetares} at the NGFP and obtain
\begin{eqnarray}
 \partial_t \zeta &=& \eta_{c} \zeta
 +\frac{25G_{\ast} \zeta}{96\pi} f(\lambda_{\ast}),\nonumber\\
f(\lambda)&=&e^{4 \lambda}\left(2 \lambda+\frac{1}{2}\right)-e^{2 \lambda}\left(\lambda+\frac{1}{2}\right)\nonumber\\
&{}&+8\lambda^2\bigl({\rm Ei}(2 \lambda)- {\rm Ei}(4 \lambda)\bigl).
\end{eqnarray}
In the present truncation involving a classical ghost kinetic term, we have
$\eta_{c}=0$. Together with the fact that $G>0$ in the physical
domain, this implies that the sign of $f(\lambda)$ decides about the sign of
the linearised flow of $\zeta$ near its Gau\ss ian fixed point. Indeed, this
function is negative for all $\lambda$, see Fig.~\ref{fig:f}.
The negative sign signals that $\zeta$ is asymptotically free. Inserting the
fixed-point values from the Einstein-Hilbert sector in the Landau-DeWitt gauge $\alpha=0$ with a
spectrally adjusted cutoff (see App. \ref{EH}), we get
$\frac{25G_{\ast}}{96\pi}f(\lambda_{\ast})=-1.404$. As long as the ghost
anomalous dimension $\eta_{c}$ remains sufficiently small, our
conclusion persists also for a larger truncation with a running ghost
kinetic term.  

Our result implies that (i) the non-Gau\ss ian fixed point in the graviton
sector is not influenced by the scalar curvature-ghost coupling, as the latter
is zero at the fixed point, and (ii) this curvature-ghost coupling is RG
relevant for the flow towards the IR. The latter property relates the initial
value of this coupling to a physical parameter that has to be fixed
by an RG condition (i.e., by an experiment). This does not necessarily imply
that the scalar curvature-ghost coupling gives rise to an {\em independent}
physical parameter. Since the background-field effective action has to satisfy
(regulator-modified) Slavnov-Taylor and background-field identities \cite{Reuter:1996cp,Reuter:1994zn,Reuter:1997gx,Pawlowski:2005xe,Gies:2006wv,Freire:2000bq}, this
operator may be related to other purely gravitonic operators. An answer to
this question requires to resolve the difference between background-metric and
fluctuation-metric dependencies which is beyond the scope of this work.

Note that our arguments straightforwardly generalise to the coupling of any
operator of the form  $\int d^d x \sqrt{\gamma} R
\cdot \mathcal{O}_{\text{s}}$, where $\mathcal{O}_{\text{s}}$ is a scalar
operator of some fields, e.g., matter fields. The interaction part of the
corresponding flow will always have a contribution $\sim G f(\lambda)$ which
supports an anti-screening flow. Of course, other contributions such as other
interaction terms, the anomalous dimensions of the matter fields, and
dimensional rescaling terms can eventually win out over the gravitational
contributions. 

\section{Conclusions}

We have contributed another building block to the asymptotic-safety scenario
for Quantum Einstein Gravity by computing the RG flow of a scalar
curvature-ghost coupling $\sim \zeta \int d^4x \sqrt{\gamma}\bar c^\mu R c_\mu$ with the aid of
the functional renormalisation group. In our present truncation involving an
Einstein-Hilbert sector, a classical ghost kinetic term, and the curvature-ghost interaction, the coupling $\zeta$ is found to be asymptotically free and
RG relevant. Therefore, it belongs to the conjectured finite set of physical
parameters which have to be fixed for an otherwise fully predictive theory of
quantum gravity. 

As this curvature-ghost coupling is marginal in a perturbative power-counting
scheme, the relevance of this coupling at the non-Gau\ss ian fixed point of
the Einstein-Hilbert system is another example of the tendency of the
fixed-point theory to increase the critical exponents of operators towards the
RG relevant regime. 

From a technical viewpoint, we have shown that a direct integration of gravity
fluctuations in the functional RG equation is possible without relying on
heat-kernel traces and propertime representations. As little is known about
heat-kernel expansions with respect to ghost operators, a running ghost sector
has not been included in an asymptotic-safety study of gravity up to
now. However, ghost operators or a more general gauge-fixing sector may carry
important pieces of information about the flow of a theory in certain gauges,
as it is the case, e.g., in Landau-gauge Yang-Mills theory in the
strong-coupling domain.  We believe that a more detailed investigation of the
ghost sector is important for a better understanding of the non-Gau\ss ian
fixed-point regime of Quantum Einstein Gravity. 

For instance, an evaluation of the ghost wave function renormalisation is
still an open question, but at the same time of primary importance as it will
both feed back into the Einstein-Hilbert sector as well as take influence on
the curvature-ghost coupling studied here.

\acknowledgments

The authors would like to thank D.F.~Litim, J.M.~Pawlowski, and F.~Saueressig
for helpful discussions. This work was supported by the DFG under contract
No. Gi 328/5-1 (Heisenberg program), FOR 723, and GK 1523.

\begin{appendix}
 \section{Einstein-Hilbert sector}\label{EH}

In this appendix, we summarise the results obtained from applying our
technique to the flow of the Einstein-Hilbert sector. As the combination of the Landau-DeWitt gauge
$\alpha=0$ together with the spectrally adjusted regulator has not been
used before, our results represent an independent verification of the
fixed-point scenario.

The spectrally adjusted flow equation has the following form in the
Einstein-Hilbert sector
\begin{widetext}
\begin{eqnarray}
 \partial_t \Gamma_k&=& \sum_{\{m \in \mathcal{M}\}} \left(\Gamma_{k\,
   m}^{(2)}\left(1+r(y) \right)\right)^{-1}\left[\partial_t \Gamma_{k\,
     m}^{(2)}r(y) + \Gamma_{k\, m}^{(2)} r'(y) \left( y(-2+\eta) +
   \frac{\partial_t \Gamma_{k\, m}^{(2)}}{Z_{\text{N}}N_{m}\bar{\kappa}^2
     k^2}\right) \right],  
\end{eqnarray}
\end{widetext}
where
\begin{equation}
 y=\frac{\Gamma_{k\, m}^{(2)}}{Z_{\text{N}}N_{m} \bar{\kappa}^2 k^2}. 
\end{equation}
Here, $\mathcal{M}$ is the set of all modes of the York decomposition of the
graviton as well as the ghost. We have also introduced a normalising factor
$N_{m}$ which is chosen such that numerical factors in front of the eigenvalue
of $-\bar{D}^2$ in the inverse propagator cancel in $y$. This guarantees that all Lanplacean momentum modes are
regularised by the same effective IR cutoff scale.  In order to regularise the
conformal instability, $N_m$ may also acquire a negative sign.

Inserting the propagators and using the Euler-MacLaurin formula for
evaluating the spectral traces, we project onto the two running couplings in
the Einstein-Hilbert sector. E.g., in four dimensions, $\partial_t
Z_{\text{N}}$ is accompanied by a factor of the curvature radius squared,
whereas the combination $\partial_t (Z_{\text{N}} \lambda)$ comes with a
factor of $r^4$. This allows for a straightforward projection and yields the
following $\beta$ functions of the dimensionless couplings:
\begin{widetext}
\begin{eqnarray}
\partial_t G&=& (2+\eta) G, \nonumber\\
  \partial_t\lambda&=&(-2+\eta) \lambda
-\frac{G}{4\pi } (5 \eta +5 e^{2 \lambda } (2 \partial_t\lambda+\eta +4 \lambda )-20 \lambda  (\partial_t\lambda+2 \lambda ) \left(\text{Ei}(2
\lambda )+\text{Log}\left(1-e^{2 \lambda }\right)-\text{Log}\left(e^{2 \lambda }-1\right)\right)\\
&{}&-10 \partial_t\lambda \text{Li}_2\left(e^{2
\lambda }\right)-20 \text{Li}_3\left(e^{2 \lambda }\right)+12 \zeta (3)+20 i \pi \lambda (\partial_t\lambda+2 \lambda )\theta(-\lambda))\nonumber\\
\eta &=& -\frac{5 G}{36 \pi } \big(2 \pi ^2+ \left(6-15 e^{2 \lambda }\right) \eta +30 (\partial_t\lambda+2
\lambda ) \text{Ei}(2 \lambda )+60 \lambda  \text{Log}\left(1-e^{2 \lambda }\right)-30 (\partial_t\lambda+2 \lambda ) \text{Log}\left(e^{2 \lambda
}-1\right)\\
&{}&+30 \text{Li}_2\left(e^{2 \lambda }\right)+30 i \pi (\partial_t\lambda+2 \lambda )\theta(-\lambda)\big).\nonumber
\end{eqnarray}
Note that branch cuts in the logarithm and the polylogarithm contribute imaginary parts which cancel for $\lambda>0$. Thereby, the flow equations are completely real for all values of $\lambda$.
%
At the fixed point, $\partial_t \lambda_k =0$ and $\eta=-2$, implying (for $\lambda>0$)
%
\begin{eqnarray}
  0&=& -4\lambda
-\frac{G}{4\pi } \left(5 e^{2 \lambda } (4 \lambda-2)-10-40 \lambda^2\left(\text{Ei}(2
\lambda )+\text{Log}\left(1-e^{2 \lambda }\right)-\text{Log}\left(e^{2 \lambda }-1\right)\right)-20 \text{Li}_3\left(e^{2 \lambda }\right)+12 \zeta (3)\right) \quad\\
-2 &=& -\frac{5 G}{18 \pi } \left(\pi ^2-6 +15 e^{2 \lambda }+30\lambda \left(\text{Ei}(2 \lambda )+\text{Log}\left(1-e^{2 \lambda }\right)-\text{Log}\left(e^{2 \lambda
}-1\right)\right)+15 \text{Li}_2\left(e^{2 \lambda }\right)\right).
\end{eqnarray}
\end{widetext}
A numerical solution yields the following fixed-point values
\begin{eqnarray}
 G_{\ast}= 0.2797,\quad 
\lambda_{\ast}=0.3407.
\end{eqnarray}
\begin{figure}[!here]
 \includegraphics[scale=0.9]{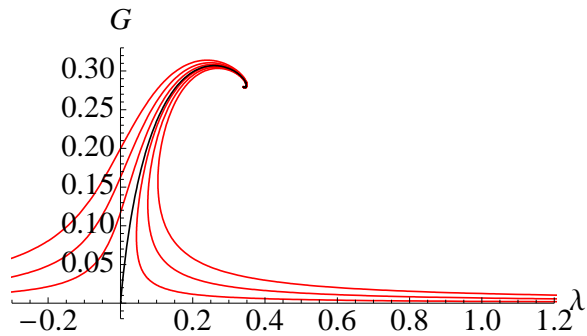}
\caption{RG trajectories emanating from the UV fixed point in the Einstein-Hilbert truncation.}
\label{flowspiral}
\end{figure}
The universal value $G_{\ast} \lambda_{\ast}=0.0953$ is very close to the
numerical values from other gauges and regulators \cite{Litim:2008tt}. The
eigenvalues of the stability matrix correspond to the critical exponents at
the fixed point (apart from a minus sign). In the present truncation, they
appear as a complex conjugate pair,
\begin{equation}
 \theta_{1,2}= 2.225 \pm i 1.572,
\end{equation}
inducing a spiral shape of the flow lines in the vicinity of the NGFP, see Fig. \ref{flowspiral}.

\end{appendix}

\end{document}